\documentclass[twocolumn,reprint, superscriptaddress, amsmath,amssymb, aps, prl]{revtex4-1}
\usepackage{graphicx}
\usepackage{amsmath}
\usepackage{color}
\usepackage{dcolumn}
\usepackage{bm}
\usepackage[
   bookmarks=true,
   colorlinks=true,
   hyperindex=true,
   citecolor=blue,
   linkcolor=blue,
   urlcolor=red
]{hyperref}

\begin{document}
\title{Probing the interaction energy of two $^{85}$Rb atoms in an optical tweezer via spin-motion coupling}

\author{Jun Zhuang}
\affiliation{State Key Laboratory of Magnetic Resonance and Atomic and Molecular Physics, Innovation Academy for
Precision Measurement Science and Technology, Chinese Academy of Sciences, Wuhan 430071, China}
\affiliation{School of Physical Sciences, University of Chinese Academy of Sciences, Beijing 100049, China}
\author{Kun-Peng Wang}
\affiliation{State Key Laboratory of Magnetic Resonance and Atomic and Molecular Physics, Innovation Academy for
Precision Measurement Science and Technology, Chinese Academy of Sciences, Wuhan 430071, China}
\author{Peng-Xiang Wang}
\affiliation{State Key Laboratory of Magnetic Resonance and Atomic and Molecular Physics, Innovation Academy for
Precision Measurement Science and Technology, Chinese Academy of Sciences, Wuhan 430071, China}
\affiliation{School of Physical Sciences, University of Chinese Academy of Sciences, Beijing 100049, China}
\author{Ming-Rui Wei}
\affiliation{State Key Laboratory of Magnetic Resonance and Atomic and Molecular Physics, Innovation Academy for
Precision Measurement Science and Technology, Chinese Academy of Sciences, Wuhan 430071, China}
\affiliation{School of Physical Sciences, University of Chinese Academy of Sciences, Beijing 100049, China}
\author{Bahtiyar Mamat}
\affiliation{State Key Laboratory of Magnetic Resonance and Atomic and Molecular Physics, Innovation Academy for
Precision Measurement Science and Technology, Chinese Academy of Sciences, Wuhan 430071, China}
\affiliation{School of Physical Sciences, University of Chinese Academy of Sciences, Beijing 100049, China}
\author{Cheng Sheng}
\affiliation{State Key Laboratory of Magnetic Resonance and Atomic and Molecular Physics, Innovation Academy for
Precision Measurement Science and Technology, Chinese Academy of Sciences, Wuhan 430071, China}
\author{Peng Xu}
\affiliation{State Key Laboratory of Magnetic Resonance and Atomic and Molecular Physics, Innovation Academy for
Precision Measurement Science and Technology, Chinese Academy of Sciences, Wuhan 430071, China}
\affiliation{Wuhan Institute of Quantum Technology, Wuhan 430206, China}
\author{Min Liu}
\affiliation{State Key Laboratory of Magnetic Resonance and Atomic and Molecular Physics, Innovation Academy for
Precision Measurement Science and Technology, Chinese Academy of Sciences, Wuhan 430071, China}
\author{Jin Wang}
\affiliation{State Key Laboratory of Magnetic Resonance and Atomic and Molecular Physics, Innovation Academy for
Precision Measurement Science and Technology, Chinese Academy of Sciences, Wuhan 430071, China}
\affiliation{Wuhan Institute of Quantum Technology, Wuhan 430206, China}
\author{Xiao-Dong He}
\email{hexd@wipm.ac.cn}
\affiliation{State Key Laboratory of Magnetic Resonance and Atomic and Molecular Physics, Innovation Academy for
Precision Measurement Science and Technology, Chinese Academy of Sciences, Wuhan 430071, China}
\affiliation{Wuhan Institute of Quantum Technology, Wuhan 430206, China}
\author{Ming-Sheng Zhan}
\email{mszhan@wipm.ac.cn}
\affiliation{State Key Laboratory of Magnetic Resonance and Atomic and Molecular Physics, Innovation Academy for
Precision Measurement Science and Technology, Chinese Academy of Sciences, Wuhan 430071, China}
\affiliation{Wuhan Institute of Quantum Technology, Wuhan 430206, China}
\date{\today}

\begin{abstract}
The inherent polarization gradients in tight optical tweezers can be used to couple the atomic spins  to the two-body motion under the action of a microwave spin-flip transition, so that such a spin-motion coupling offers an important control knob on the motional states of optically trapped two colliding atoms. Here, after preparing two elastically scattering $^{85}$Rb atoms in the three-dimensional ground-state in the optical tweezer, we employed this control in order to probe the colliding energies of elastic and inelastic channels. The combination of microwave spectra and corresponding s-wave pseudopotential model allows us to infer the effect of the state-dependent trapping potentials on the elastics colliding energies, as well as to reveal how the presence of inelastic interactions affects elastic part of the relative potential. Our work shows that the spin-motion coupling in a tight optical tweezer expand the experimental toolbox for fundamental studies of ultracold collisions in the two body systems with reactive collisions, and potentially for that of  more complex interactions, such as optically trapped atom-molecule and molecule-molecule interactions.

\end{abstract}
\maketitle

\section{I. introduction}
\label{s1}

Since the successful  loading of single atoms from a cold atomic ensemble into microscopic optical dipole trap--namely  optical tweezer~\cite{1Schlosser2001}, this approach has progressed to having a profound impact on many research areas through bottom-up scaling with an unprecedented level of programmability and scalability~\cite{Kim2016,Endres2016,Barredo2016,Barredo2018,Kumar2018,Sheng2022,Singh2022}, ranging from quantum simulations of many-body physics~\cite{9Browaeys2020,Zoller2023}, quantum computing~\cite{Saffman2016,Sheng2018,Bluvstein2022,Graham2022,8Evered2023,10Bluvstein2023}, metrology~\cite{Madjarov2019,Covey2019,Norcia2019,Young2020}, ultracold collisional physics~\cite{Andersen2022} and association of single molecules~\cite{Liu2018,12HeXD2020} and array of single molecules~\cite{Anderegg2019,Zhang2022}.

Particularly, the high-level internal states control and single-particle level detection allow ones to build an extremely clean platform for the study of ultracold collisions~\cite{11XuP2015,Sompet2019,Reynolds2020,Cheuk2020,Hood2020,Weyland2021,Brooks2022}. Beyond these advances, the individual pairs of atoms in optical tweezers has been transferred to weakly bound molecules via Raman transitions~\cite{Liu2019,Yu2021} and magneto association~\cite{Zhang2020}. Interestingly, the optical tweezer itself has been proved to offer new manners to molecular association without using Fano-Feshbach resonances so as to allow a wider range of molecular species. For example, we can use methods of coupling two atoms' relative motion and spins~\cite{12HeXD2020} and of merging optical tweezers~\cite{Ruttley2023}.

To provide a sufficiently strong trapping potential for atomic trapping, the beam waist of the optical tweezers used in current experiments is typically comparable to the laser wavelength. This leads to the presence of a longitudinal electric field near the focus point, inducing a polarization gradient~\cite{15Richards1959}. This polarization gradient creates an equivalent gradient magnetic field for the trapped atoms\cite{27RamancoolingPRX,23Thompson2013}, allowing for the realization of single-atom spin-motional coupling~\cite{16Dareau2018} and high-precision manipulation of motional states~\cite{17WangKP2020}. In the case of two-atom systems, the coupling between the spins of the two atoms and their relative motion or center-of-mass motion (spin-motion coupling, SMC) can be straightforward achieved. Under this coupling regime, it has been experimentally demonstrated that an internal state transition of one atom in the two-atom system, driven by a microwave (MW) pulse, can induce transitions between different eigenenergies of the two-atom system. This enables precise measurement and control of two-atom interactions. This effect has been utilized in the study of $^{85}$Rb-$^{87}$Rb interaction and single molecule synthesis~\cite{12HeXD2020}.The scattering length of $^{85}$Rb-$^{87}$Rb is positive, resulting in a repulsive potential for their interaction. It would be interesting to apply SMC method to study the ultracold collisions of atoms with negative scattering length. The corresponding interaction in an optical tweezer becomes attractive, giving rise to a bound state with a ground state energy lower than the zero point energy of a harmonic oscillator~\cite{19Busch1998}. Specifically, the $^{85}$Rb atoms have large negative background scattering lengths and large difference in scattering lengths between singlet and triplet electronic states, which leads to large hyperfine-exchange collision rates~\cite{18Claussen2003}. Therefore, two ultracold $^{85}$Rb atoms in an optical tweezer lends itself well to test contact pseudopotential model with complex scattering lengths~\cite{28Idziaszek2016}.

In this work, we utilize the SMC to study the scattering properties of individual pairs of $^{85}$Rb atoms prepared in the 3D ground state of an optical tweezer. From the resulting MW spectra, that are resonant transitions in the motional state manifold under the action of a MW spin-flip transition, the  collision energies of elastic channel \{$| F,m_F\rangle = |3,-3\rangle$ + $|2,-2\rangle$\}(abbreviated as \{3,-3;2,-2\}) and inelastic channel \{3,-3;3,-1\}  are deduced respectively. Given the known values of scattering length, the pseudopotential model calculations enable us to confirm that the attractive interaction energies of the \{3,-3;2,-2\} channel is lifted up due to the state-dependent trapping potentials. However, for the inelastic \{3,-3;3,-1\} channel, the deduced interaction energy is found to be obviously smaller than the calculated ones according to the pseudopotential model calculations with given complex scattering length for the reactive collision. This discrepancy reveal that the  pseudopotential model may be inapplicable to the prediction of atom-atom interaction energies in the presence of inelastic decay, and so that  a more realistic atom-atom interaction calculation is needed. This study highlights the importance of SMC approach for the fundamental study of ultracold collisions.

This paper is organized as follows. In Sec. II, we describe the associated experimental setup. In Sec. III, we detail the preparation of two $^{85}$Rb atoms in the three-dimensional ground-state in an optical tweezer. In Sec. IV, we present the two-atom quantum motion resolved the MW spectra and the extraction of  collision energies, and the analysis in the framework of  s-wave pseudopotential model. In Sec. V, we present conclusions and outlook the promising applications of SMC scheme.


\section{II. experimental apparatus for ultracold atom}

\begin{figure*}[htbp]
\centering
\includegraphics[width=18cm,height=9cm]{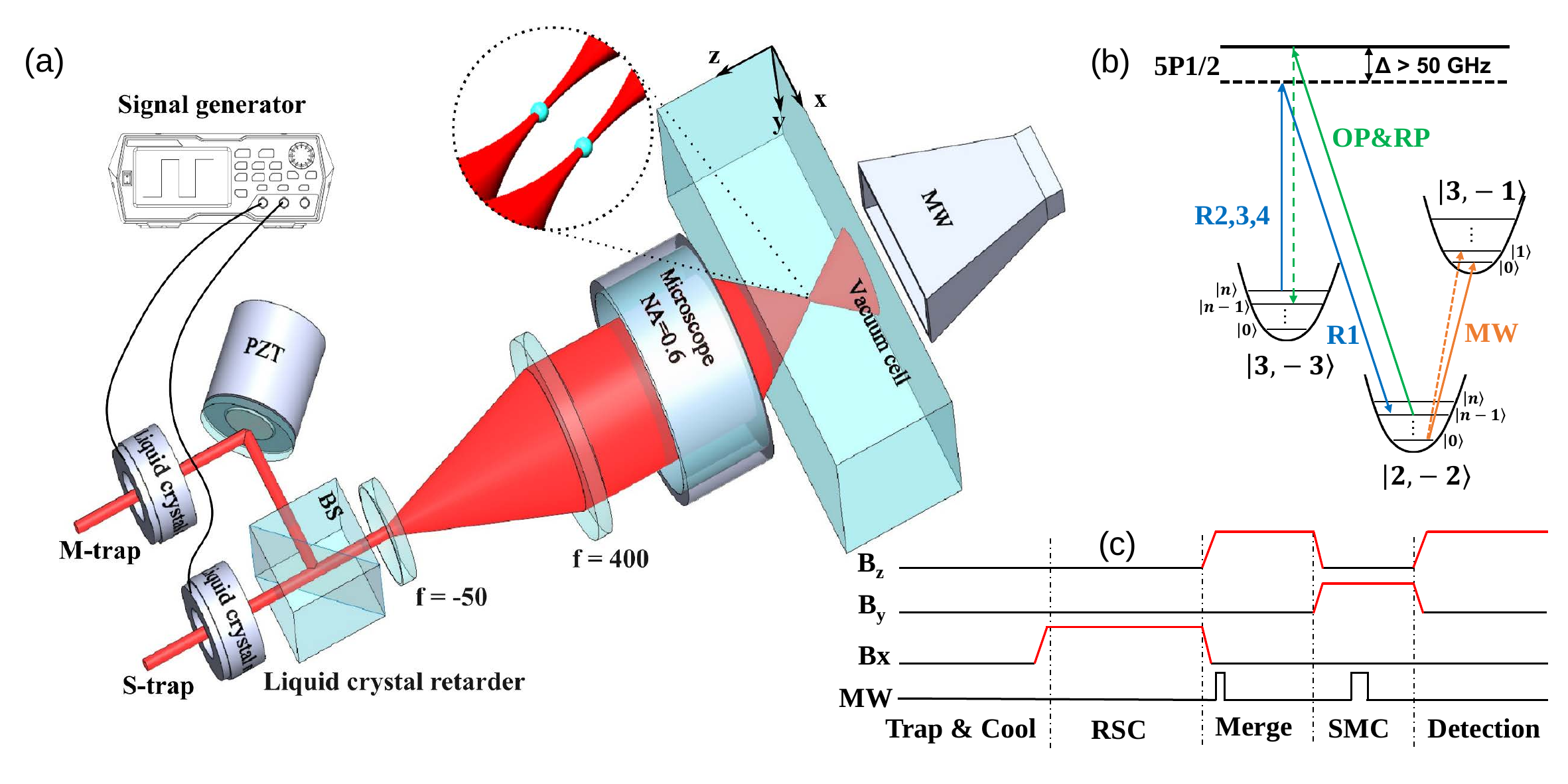}
\caption{Schematic diagram of experimental setup and the energy level transition of $^{85}$Rb. (a) Schematic diagram of the experimental setup. Two 852-nm beams are combined by  beam splitter (BS), then expanded 8 times by a beam-expander group (lens with f = -50 mm and f = 400 mm), then enter into a microscope with NA = 0.6 to obtain two strongly focused optical tweezers with the waists of 0.75 $\mu$m; The liquid crystal retarders are employed to adjust the polarization of dipole optical light; The PZT is used to precisely adjust the pointing of M trap; The MW horn is used to emit the MW pulse for MW spectra. (b) Schematic diagram of the energy level transition of $^{85}$Rb. (c) Schematic diagram of the experimental timing sequence. }
\label{fig:fig1}
\end{figure*}

Fig.~\ref{fig:fig1}(a) is the schematic diagram of experiment setup. The experimental arrangement employed in this study is similar to our prior works~\cite{22WangKP2019,12HeXD2020}. To attain independent control over two ultracold $^{85}$Rb atoms, we engineer two distinct collimated 852-nm trapping beams, each with an approximate diameter of 3 mm. These beams are denoted as static trap (S-trap) and movable trap (M-trap) respectively. The two laser beams are combined via a polarization-independent beam splitter (BS), then expanded 8 times by a beam expander group (lens with f = -50 mm and f = 400 mm), and lastly focused via a 0.6 numerical-aperture (NA) objective. The resulting two focused waists are about  0.75 $\mu$m with a spacing of 4 $\mu$m. To ensure precise control over the polarization of the optical tweezers on demand, a pair of liquid crystal retarders (LCRs, LCB1111-B from Thorlabs) are separately employed to dynamically adjust the polarizations of the S-trap and M-trap. A piezoelectric (PZT) transducer is used to adjust the pointing of M-trap to precisely merge and split two atoms.

Fig.~\ref{fig:fig1}(b) presents a schematic representation of the energy level transitions of the $^{85}$Rb atom. The hyperfine state of $| F,m_F\rangle = |3,-3\rangle$ and $|2,-2\rangle$ are relevant to Raman sideband cooling (RSC) and state-dependent transfer. The MW pulses emitted by a horn is used to drive the transition of $|2,-2\rangle \rightarrow |3,-1\rangle$ for MW spectra. Fig.~\ref{fig:fig1}(c) shows the schematic diagram of the experimental timing sequences. The optical tweezers are horizontally polarized in RSC and SMC, and the magnetic field  is correspondingly set along the $x$ direction (B$_{x}$) to suppress the polarization gradient effect during RSC period, whereas the magnetic field is needed to be oriented along the $y$ direction  (B$_{y}$) to turn on the polarization gradient effect during SMC period.  In the state-dependent merging and detection process, the magnetic field is set in the z direction (B$_{z}$) in order to construct state-dependent potential traps.


\section{III. preparation of a pair of ultracold $^{85}$Rb atoms}

\begin{figure}[htbp]
\centering
\includegraphics[width=8.6cm]{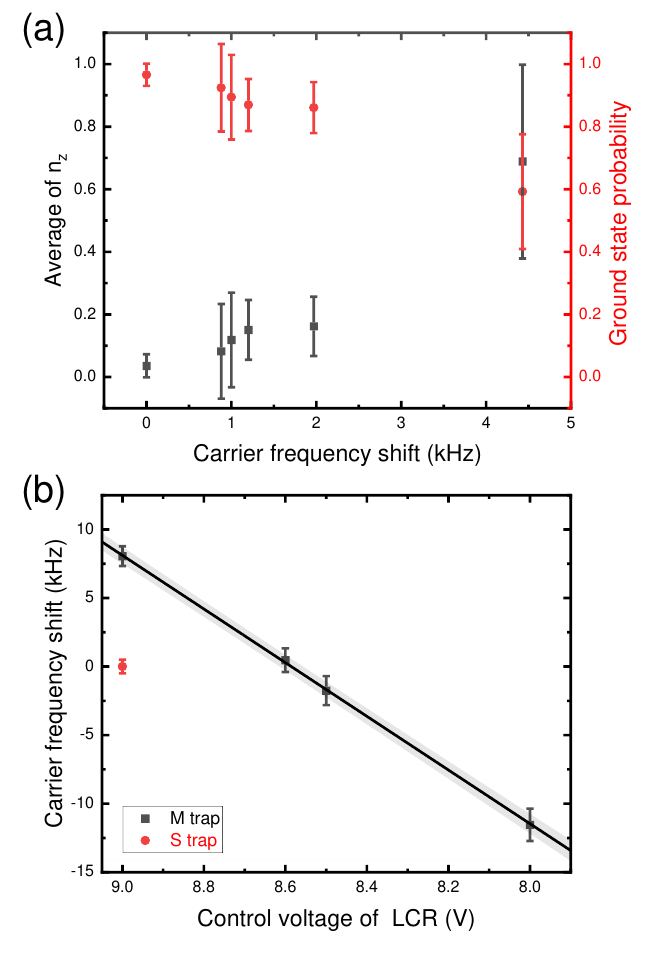}
\caption{The effect of Raman transition carrier frequency shift on axial cooling, and the relationship between the carrier frequency shift and control voltage of LCR. (a) The relationship between RSC carrier frequency shift and the average quantum number and the ground state preparation probability of axial direction. When the carrier frequency shift is greater than 4.4 kHz, the average axial quantum number increases to above 0.7(3), and the ground state preparation probability decreases to below 0.6(2). (b) The relationship between the carrier frequency shift and and control voltage of LCR. }
\label{fig:fig2}
\end{figure}

To achieve the preparation of a pair of $^{85}$Rb ultracold ground state atoms, the first step is to achieve RSC cooling in the 3D ground state. In the RSC process, the Raman transition carrier frequencies of the two atoms need to be calibrated. To reduce the vector light shift (VLS) of the two atoms, we finely adjust the polarization of both traps to horizontal linear polarization using a polarization analyzer (SK010PA-NIR). However, there is still a frequency difference of about 8 kHz between the carrier transitions of the two atoms. In the RSC process, the axial sideband cooling is the most challenging step due to the relatively small trapping frequency (2$\pi \times 25$ kHz) in the axial direction. The axial Rabi frequency is sensitive to separate the carrier peak and sideband peaks of the Raman transition because of the small trapping frequency. A large Rabi frequency can cause heating due to off-resonant transitions, while a too small Rabi frequency can lead to diminished transition efficiency and decelerated cooling rates~\cite{27RamancoolingPRX,22WangKP2019}. Therefore, in order to balance this, the axial Rabi frequency is typically set to around 6 kHz, and the Rabi frequency of the $\Delta n = -1$ sideband is only about 2 kHz. Consequently, in the RSC process, the carrier transition frequencies of the two atoms in S-trap and M-trap need to be as consistent as possible. Fig.~\ref{fig:fig2}(a) shows the influence of the transition carrier frequency shift (CFS) on the axial sideband cooling for single $^{85}$Rb atom. When the carrier frequency shift is greater than 4.4 kHz, the average axial quantum number increases to above 0.7(3), and the preparation probability of the axial ground state decreases to below 0.6(2).

To precisely adjust the consistency of the transition frequencies of the atoms, the relationship between the transition carrier frequency of the atom in M-trap and the control voltage of the LCR is measured as shown in Fig.~\ref{fig:fig2}(b). We can accurately set the control voltage of the LCR to make the carrier transition frequencies of the two atoms consistent. After calibrating the Raman transition carrier frequencies, we successfully achieve Raman sideband cooling of the two atoms and obtain a pair of $|3,-3\rangle$ atoms in the 3D ground state. After cooling, the average quantum numbers $\{{\overline{n}_x,\overline{n}_y,\overline{n}_z}\}$ for atoms in S-trap and M-trap are $\{{0.03(5),0.02(3),0.04(4)}\}$ and $\{{0.02(4),0.04(4),0.03(3)}\}$ respectively. The final probability of the 3D ground state is determined to be $0.91(6)$ and $0.90(6)$ for the atom in the S-trap and M-trap, respectively.

The second step is to implement non-heating merging of the two atoms, which relies on using state-dependent potentials via utilizing the the vector light shifts (VLSs)~\cite{22WangKP2019}. To do so, we first switch the magnetic field to B$_{z}$ and adjust the linearly polarized S-trap (M-trap) to a $\sigma^{+}$ ($\sigma^{-}$) one by dynamically control LCRs. As a result, influenced by the VLS, the difference in the frequency of $|3,-3\rangle \rightarrow |2,-2\rangle$ transition between the atom in the S-trap and the one in the M-trap is high up to level of MHz. Such a frequency gap  allows us to use MW pulse to selectively drive the $|3,-3\rangle \rightarrow |2,-2\rangle$ transition for the atom in M-trap with extremely low cross talk. The atom in M-trap is then moved into the S-trap by PZT, achieving non-heating merging without changing the quantum numbers of the two atoms motion. Finally, the dipole light of M-trap is turned off adiabatically, and a pair of $^{85}$Rb ultracold atoms in 3D ground state, each in a different hyperfine magnetic sublevel ($|3,-3\rangle$ and $|2,-2\rangle$) and with completely controllable internal and external states, is prepared in an optical tweezer, such a spin combination is stable against hyperfine changing spin collisions.

\section{IV. two-atom motion resolved-microwave spectra and Analysis}

\begin{figure}[htbp]
\centering
\includegraphics[width=9cm]{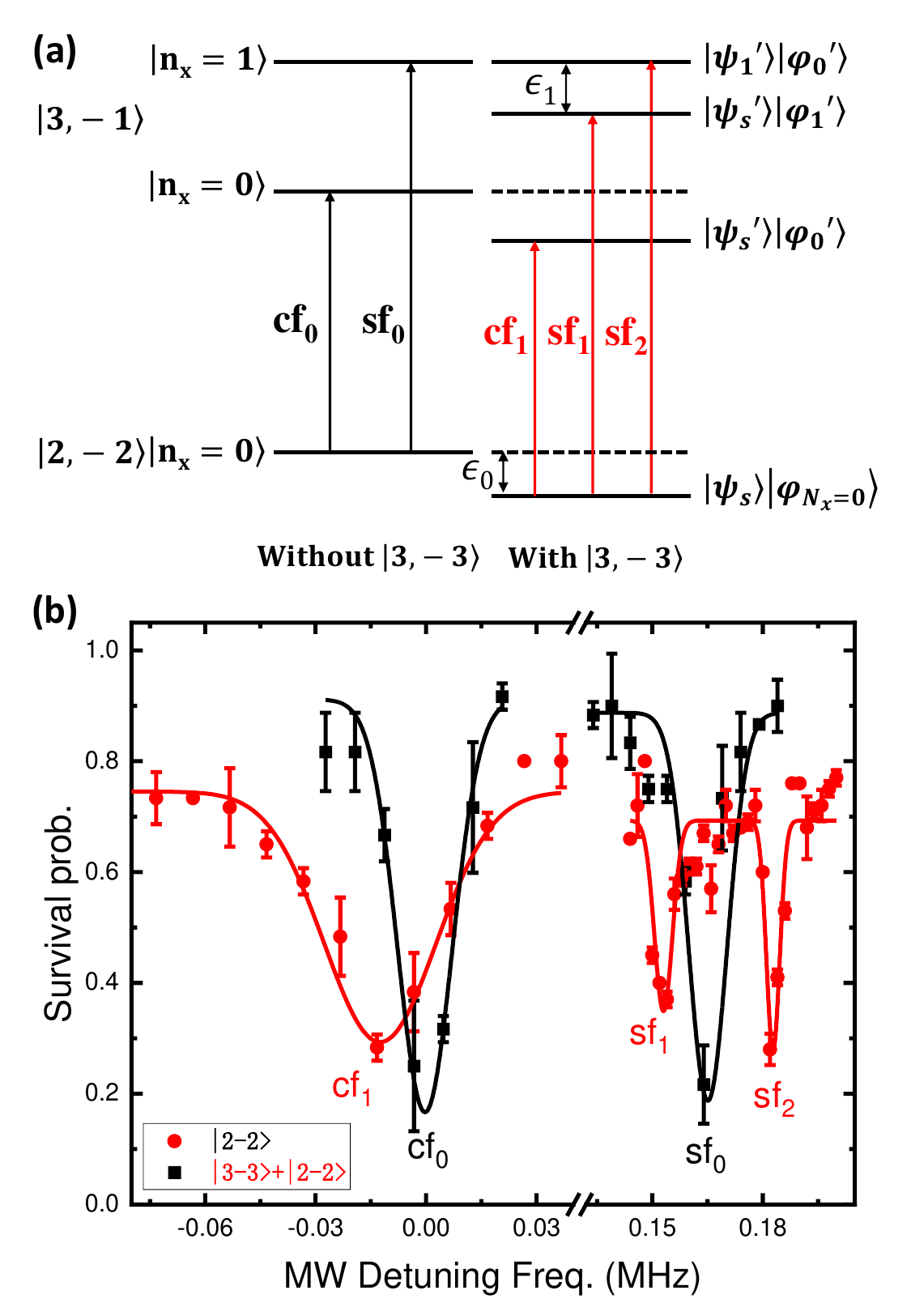}
\caption{Schematic diagram of the transition and the MW spectra via SMC. (a) Schematic diagram of the vibrational transition of $^{85}$Rb atom. (b) Single atomic MW spectra of $|2,-2\rangle$ $\rightarrow$ $||3,-1\rangle$ (black) and diatomic transition spectra of $|3,-3\rangle|2,-2\rangle \rightarrow |3,-3\rangle|3,-1\rangle$ (red). The spectra set the single atom transition carrier frequency (3028.0050(8) MHz) as the reference, and the solid curves are Gaussian fits of the data. The detuning frequency of the single-atom carrier peak is $cf_0 = 0.0(8)$ kHz, while the detuning frequency of the diatomic carrier peak is $cf_1 = -10.0(9)$ kHz. Furthermore, the detuning frequency of the single-atom sideband peak is measured to be $sf_0 = 163.2(5)$ kHz. Additionally, the diatomic system exhibited two distinct peaks: the center of mass motion peak at $sf_1 = 153.0(5)$ kHz and the relative motion peak at $sf_2 = 182.8(3)$ kHz. }
\label{fig:fig3}
\end{figure}

Having prepared a pair of ultracold $^{85}$Rb atoms in an optical tweezer, we now describe the study of two-atom MW spectra in the presence of SMC. Fig.~\ref{fig:fig3}(a) depicts the vibrational transition diagram of $|2,-2\rangle$ $\rightarrow$ $|3,-1\rangle$ with and without the atom in $| 3,-3\rangle$. For the harmonic trapping potential, the center-of-mass (c.m.) and relative motion of two colliding homonuclear atoms are decoupled, thereby rendering the two-atom transition equivalent to $|\psi_s \rangle| \varphi_{N_x = 0} \rangle \rightarrow | \psi_s' \rangle | \varphi_{N_x} '\rangle $ ($| \psi_s \rangle$ and $| \varphi_r \rangle$ represent the c.m. and relative motional states respectively; \{$N_x = {0,1,\text{...}}$\} denotes the quantum number of the c.m. motion in the x direction); and the subscript (s) denotes the ground state of relation motion. In the ultracold two-atom regime, the scattering is purely of s-wave character and  so that the exact interatomic potential is conventionally represented by the well-known $\delta$-function pseudopotential model. When the atoms in the relative motion states with odd quantum numbers do not feel the interaction because the relative wave function is zero at the $\delta$ function~\cite{19Busch1998,24Idziaszek2006}, meaning that the corresponding wavefunctions are just harmonic oscillator ones. For example, when the atoms occupy the first excited state of the relative motion $| \psi_1'\rangle$ in the x direction, the relative motion energy is equal  to the first excited motional state of single atoms. Due to the attractive interactions predetermined by negative scattering lengths, the energy level $| \psi_s\rangle| \varphi_{N_x=0}\rangle$ are consequently shifted lower. Here, we use  $\epsilon_0$ and $\epsilon_1$ to denote the interaction energy of channels of \{3,-3;2,-2\} and \{3,-3;3,-1\} in the ground states of relative motion, respectively. For the single atom, there exists only one $\Delta n_x = 1$ sideband transition $sf_0$. However, those interaction potential will induce a splitting of the two-atom sideband transition into c.m. motional transition ($sf_1$) and relative motional transition ($sf_2$). Thus, the difference between $sf_0$ and $sf_2$ is equal to the interaction energy of the channel \{3,-3;2,-2\} (i.e. $\epsilon_0 = sf_2 - sf_0$), and the spacing between $sf_1$ and $sf_2$ is equal to the interaction energy of \{3,-3;3,-1\} (i.e. $\epsilon_1 = sf_2 - sf_1$).

To record the two-atom MW spectra, the polarization of optical tweezer laser and the magnetic field are respectively changed to x and  y direction so as to turn on the two atom SMC.  We subsequently record two-atom MW spectra by applying rectangular shape pulses to drive the hyperfine transition $| 2,-2\rangle \rightarrow | 3,-1\rangle$. The outcome two atoms, $| 3,-1\rangle$ and $| 3,-3\rangle$, have vector light shifts of the same sign and move together during the species-dependent transport, leading to the disappearance of the atomic fluorescent signals. The resulting spectra with three interaction-shifted peaks are shown in Fig.~\ref{fig:fig3}(b), in which the carrier and the sideband transitions for the single atoms are also plot for comparison. From left to right, the two atom peaks are label by $\{cf_1,sf_1,sf_2\}$, respectively. For this measurements, the depth of the dipole trap in the experiment is approximately 1.6 mK, and the oscillation frequency of trapped atoms are about 164 kHz and 25 kHz in the radial and axial directions respectively.  And the magnetic field intensity is approximately 5.52 G.

The peak $cf_1$ is of the resonant transitions between the motional ground states together with the spin-flip transition  $| 2,-2\rangle \rightarrow | 3,-1\rangle$. The shift with respective to the carrier of single atoms gives the difference of interaction energies between the \{3,-3;2,-2\} channel and \{3,-3;3,-1\} channel. The peak $cf_1$ is similar to the one presented in the previous work of Raman spectroscopy of two atoms (Na-Cs) in an optical tweezer~\cite{Hood2020}. The spacing between the peaks $cf_1$ and  $sf_1$ is equal to the radial trapping frequency ($sf_1 - cf_1 = sf_0 - cf_0$), so that it is identified as the spin-flip transition together with the motional transition $N_x = 0\rightarrow N_x = 1$ in the c.m. motion. The peak $sf_2$ corresponds to the transition $| \psi_s\rangle| \varphi_{N_x=0}\rangle \rightarrow | \psi_1'\rangle  | \varphi_0'\rangle$, where $ | \psi_1'\rangle$ denotes the the first excited state of the relative motion.

\begin{figure}[htbp]
\centering
\includegraphics[width=9cm]{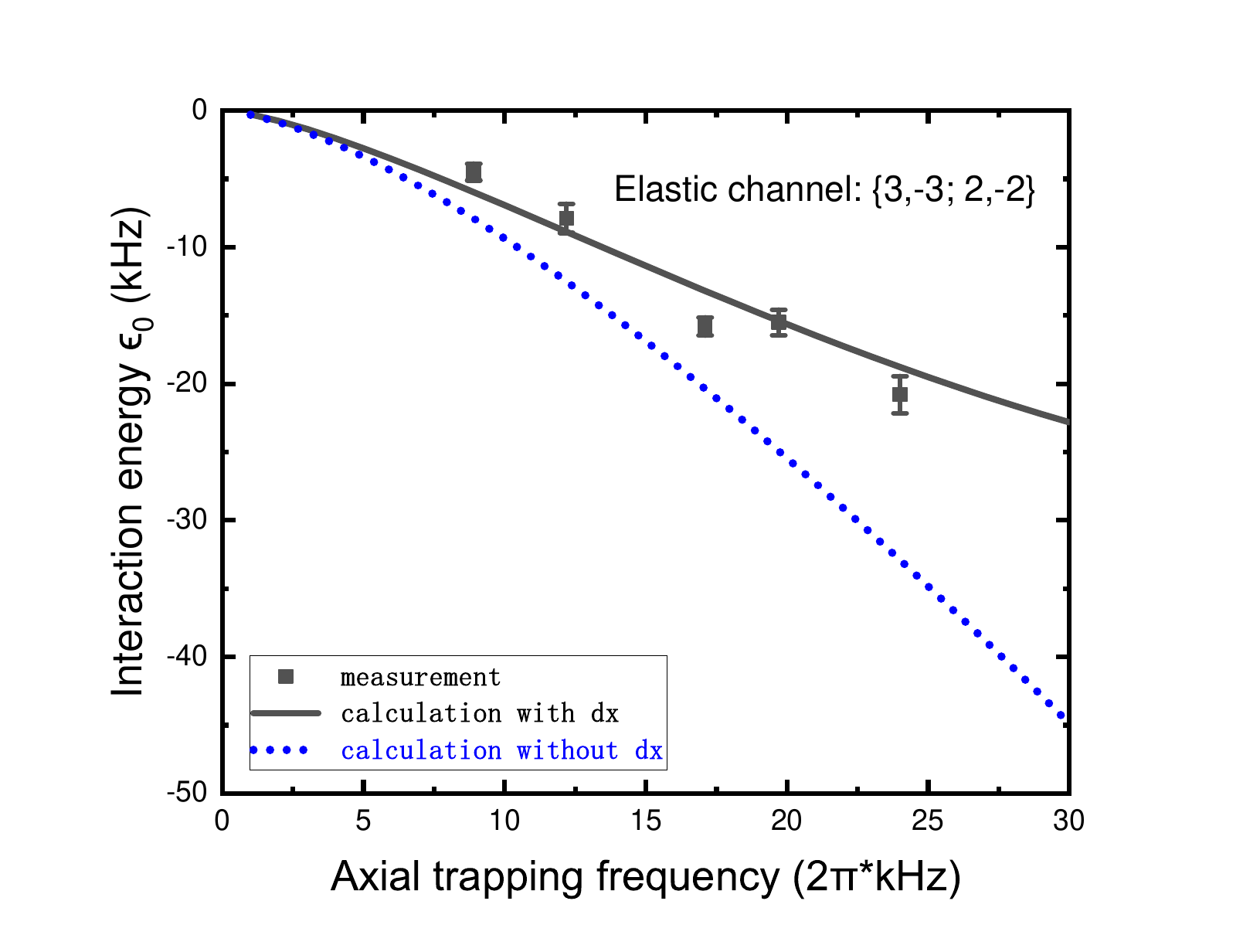}
\caption{The relationship between the interaction energy of the initial state ($|2,-2\rangle|3,-3\rangle| \psi_s\rangle|\varphi_0\rangle$) and the axial resonant frequency of the optical tweezer. The black squares represent the measured values. The accompanying error bars are statistic standard deviation for the average. And the solid black line and the dashed blue line represent the theoretical calculation of Eq.(1) values with and without the correction for the perturbation of $dx$ respectively.}
\label{fig:fig4}
\end{figure}

After the spectral  identification, the extraction of  interaction energy of specific channel  is straightforward.  The measured interaction energies of elastic channel  \{3,-3;2,-2\} are
are plotted in the Fig.~\ref{fig:fig4} as a function of axial trapping frequencies. To understand the experimental results, we adopt the analytical results for the pseudopotential model
in a cylindrically symmetric harmonic trap~\cite{24Idziaszek2006}. Briefly, the eigenenergies are calculated by the roots of the following equation with given parameters :

\begin{eqnarray}\label{eq1}
\frac{dz}{a_s} & = & \frac{2  \Gamma (\epsilon)}{\Gamma (\epsilon-\frac{1}{2})}-\frac{\Gamma (\epsilon) \sum \limits_{m=1}^{\eta -1} F_1(1,\epsilon,\epsilon+\frac{1}{2},e^{\frac{i 2 \pi  m}{\eta} })}{\Gamma (\epsilon+\frac{1}{2})}
\end{eqnarray}
Here, $a_s$ is the s-wave scattering length, m is the angular quantum number, $\epsilon$ is the interaction energy, and $\eta = \omega_\perp/\omega_z$ is the ratio of the radial harmonic frequency to the axial harmonic frequency, $\Gamma(j)$ is the gamma function, $F1(a,b,c,j)$ is the hypergeometric function, $dz = \sqrt{ \hbar /(\mu \omega_z)}$ ($\omega_z$ is the axial harmonic frequency), and $\mu = m_{85}/2$ is the reduced mass of two $^{85}$Rb atoms. By employing the coupled channel theory~\cite{26Hutson2019}, we can calculate that the scattering length of the \{3,-3;2,-2\} channel as $a_i = - 391$ $a_0$ ($a_0$ is the Bohr radius). In our system, we need to take the effect of the polarization gradient on the $|3,-3\rangle$ and $|2,-2\rangle$ states into account. We note that the $|3,-3\rangle$ and $|2,-2\rangle$ state have a relative the wavefunction displacement $dx \approx 18.0$ nm along the $x$ direction in this experiment.  According to Ref.~\cite{25Krych2009}, this displacement induces a variation in the interaction energy perturbation,  denoted as $\epsilon'$. Its magnitude is approximately $\epsilon' \approx \mu (dx)^2 (\eta \omega _z)^2 /(2\hbar )$.  The results of calculation are plotted in the Fig.~\ref{fig:fig4}, in which the black solid line and blue dashed one represent the theoretical calculation values with and without correction for the perturbation $\epsilon'$, respectively. Notably, the experimental values match well with the theoretical calculation results after incorporating the perturbation of $\epsilon'$.

\begin{figure}[htbp]
\centering
\includegraphics[width=9cm]{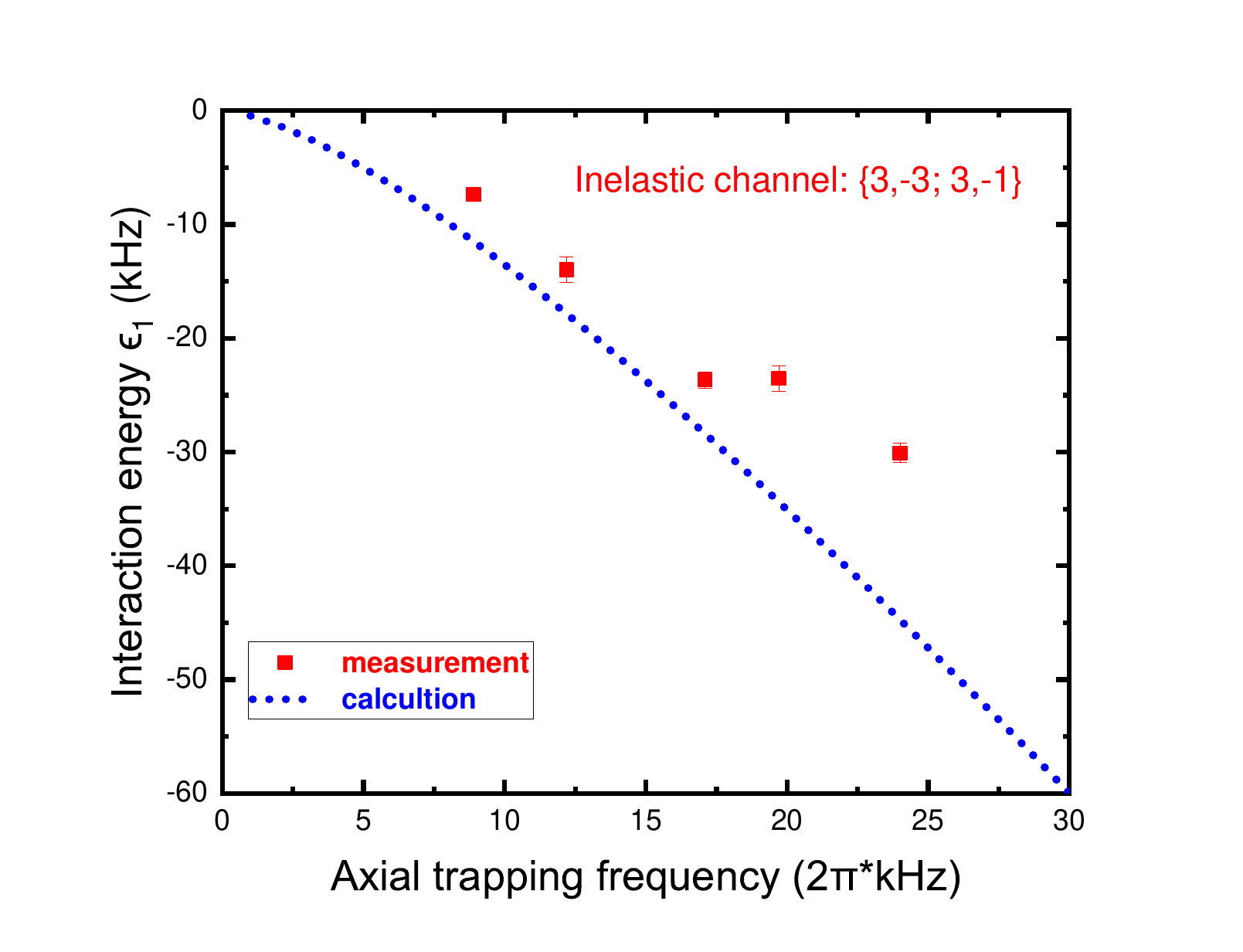}
\caption{The relationship between the interaction energy of the inelastic channel \{3,-3;3,-1\} and the axial trapping frequencies. The red squares represent the measured values. The accompanying error bars are statistic standard deviation for the average. The dashed blue line represents the predictions of the energy by applying the contact pseudopotential with complex scattering length. }
\label{fig:fig5}
\end{figure}

In the end, we'll discuss the behavior of interaction energies of the inelastic channel \{3,-3;3,-1\} in an optical tweezer. For this channel, inelastic hyperfine changing collision occurs and the atomic interaction can be modeled with complex scattering length $a_{inelastic} =\alpha- i\beta $, where the imaginary part $\beta$ is responsible for inelastic spin relaxation from the entrance channel~\cite{Hutson2007}.  Recently, by applying the contact pseudopotential with complex scattering length to a system of two ultracold particles confined in a spherically symmetric harmonic trap, the properties of eigenenergies and eigenfunctions as a function of the real and imaginary parts of the scattering length has been theoretically investigated~\cite{28Idziaszek2016}. Here, we follow this work to calculate the interaction energies with replacing the scattering length with the complex scattering length in the Eq.(1). The real roots of Eq.(1) are of interaction energies. Specifically, the values of $\alpha$ and $\beta$ are - 596$a_0$ and -43$a_0$, respectively, that are also calculated from the coupled-channel calculation program~\cite{26Hutson2019}. The predictions of interaction energies are plotted as a function of axial trapping frequencies in the Fig.~\ref{fig:fig5} , see the dashed curve. Compared with measured data, the theoretical energies are obviously larger than the experimentally measured ones. This discrepancy suggests that the contact pseudopotential with complex scattering length is too simplifier to capture the inelastic collision of two ultracold atoms confined in an optical trap, and also qualitatively reveal how the presence of inelastic interactions affects elastic part of the relative potential.


\section{V. conclusions}


In conclusion, we have successfully prepared a pair of $^{85}$Rb atoms in the three-dimensional ground-state through sequentially implementing RSC and species-dependent transport techniques, and then recorded the relative and center-of-mass motion of resolved MW spectra by taking advantage of SMC in an optical tweezer. Combing the resulting MW spectra and corresponding s-wave pseudopotential model, we have evaluated the effect of the external confinement on the collision energy of the elastic channel. Furthermore, we have found out that the measured collision energy of inelastic channel is smaller than the one determined by the real part of the complex scattering length, confirming the need of further investigation of the relative theory for calculating the collision energy for the inelastic channel. The SMC method can also be applied to the atom-molecule and molecule-molecule systems in optical tweezers. The realization of ultracold two $^{85}$Rb atoms reservoir is also an important step towards making a single molecule and studying coherent spin-mixing dynamics.

\section{acknowledgments}


This work was supported by the National Natural Science Foundation of China under grants No. 12122412, No. U22A20257, No. 12121004, No. 12241410, No. 12004395, and No. 12104464, the Project for Young Scientists in Basic Research of CAS under grant No. YSBR-058, the Key Research Program of Frontier Science of CAS under grant No. ZDBSLY-SLH012, the National Key Research and Development Program of China under Grant No. 2021YFA1402001, the Major Program (JD) of Hubei Province under Grant No. 2023BAA020, and the Natural Science Foundation of Hubei Province under Grant No. 2021CFA027.



\end{document}